\documentclass{camera}
\usepackage{graphicx}  % uncomment this if your want to insert figures

\begin{document}

%%%%%%%%%%%%%%%%%%%%%%%%%%%%%%%%%%%%%%%%%%%%%%%%%%%%%%%%
% The title, only the first letter capitalized; if you want to split it in
% two or more lines, put a \\ macro at each line break
% example: 
%   \title{Title: first line\\ second line}
%
\title{Neutron-proton elliptic flow in Au + Au}

%%%%%%%%%%%%%%%%%%%%%%%%%%%%%%%%%%%%%%%%%%%%%%%%%%%%%%%%
% The author(s), separated by commas; do not put a
% comma before the last author, use instead the \and
% macro which produces a normal ``and'' in the
% caps/small caps context
%
\author{W.\ Trautmann,$^{1}$ M.\ Chartier,$^2$ Y.\ Leifels,$^1$ 
R.C.\ Lemmon,$^{3}$\\ 
Q.\ Li,$^4$ J.\ {\L}ukasik,$^5$ A.\ Pagano,$^6$ P.\ Paw{\l}owski,$^5$ 
P.\ Russotto$^7$\\
\and P.Z.\ Wu$^2$}

%%%%%%%%%%%%%%%%%%%%%%%%%%%%%%%%%%%%%%%%%%%%%%%%%%%%%%%%
%
\organization{
$^{1}$ GSI Helmholtzzentrum 
%f\"{u}r Schwerionenforschung 
GmbH, D-64291 Darmstadt, Germany\\
$^{2}$ University of Liverpool, Liverpool L69 7ZE United Kingdom\\
$^{3}$ STFC Daresbury Laboratory, Warrington, WA4 4AD United Kingdom\\
$^{4}$ School of Science, Huzhou Teachers College, Huzhou 313000, China\\
%$^{4}$ FIAS, Universit\"{a}t Frankfurt, D-60438 Frankfurt am Main, Germany\\
$^{5}$ IFJ-PAN, Pl-31342 Krak\'ow, Poland\\
$^{6}$ INFN-Sezione di Catania, I-95123 Catania, Italy\\
$^{7}$ INFN-LNS and Universit\`{a} di Catania, I-95123 Catania, Italy
}

\maketitle

\begin{abstract}
The elliptic flow of neutrons, protons and light complex particles in
reactions of neutron-rich systems at relativistic energies is proposed as an 
observable sensitive to the strength of the symmetry term in the equation of state 
at supra-normal densities. Preliminary results from a study of the existing 
FOPI/LAND data for $^{197}$Au + $^{197}$Au collisions at 400 A MeV with the UrQMD 
model favor a moderately soft symmetry term with a density 
dependence of the potential term proportional to $(\rho/\rho_0)^{\gamma}$ 
with $\gamma \approx 0.9 \pm 0.3$.
\end{abstract}

%%%%%%%%%%%%%%%%%%%%%%%%%%%%%%%%%%%%%%%%%%%%%%%%%%%%%%%%
% Write the text starting from here and using the usual
% LaTeX commands.
%
Considerable efforts are presently underway in order to determine 
the equation of state of asymmetric nuclear matter which is of fundamental 
importance to both nuclear physics and astrophysics~\cite{lipr08}. 
While fairly consistent constraints for the symmetry energy near normal nuclear matter 
density have been deduced from recent data~[1-3],
%\cite{lipr08,klimk07,tsang09}, 
much more work is still needed to probe its high-density behaviour. 
This requires reaction studies at sufficiently high energies and suitable probes 
sensitive to mean-field effects in the initial compressed stage of the reaction as, 
for example, the neutron-proton differential transverse and elliptic 
flows~\cite{li02,greco03}.

In two experiments at GSI combining the LAND and FOPI (Phase 1) detectors, 
both neutron and 
hydrogen collective flow observables from $^{197}$Au + $^{197}$Au collisions at 
400, 600 and 800 A MeV have been measured~\cite{leif93}. This data set is 
presently being reanalyzed in order to determine optimum conditions 
for a dedicated new experiment, but also with the aim to produce constraints
for the symmetry energy by comparing with predictions of state-of-the-art 
transport models. Here, we report first results obtained with the UrQMD model which 
has recently been adapted to heavy ion reactions at intermediate energies~\cite{qli05}.

\begin{figure}[htb!]
\centering
\vskip -0.1cm

\includegraphics*[height=7.7cm]{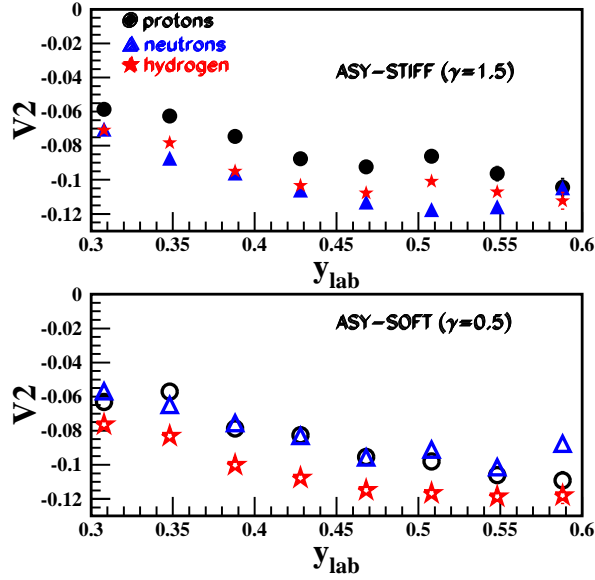}

\caption{Elliptic flow parameter $v_2$ for mid-peripheral $^{197}$Au + $^{197}$Au 
collisions at 400 MeV per nucleon as calculated with the UrQMD model for protons (circles), neutrons (triangles), and the total hydrogen yield (stars) as a function of the laboratory rapidity $y_{\rm lab}$. The results have been filtered to correspond to the geometrical acceptance of the LAND setup used in the joined experiment. The predictions obtained with a stiff and a soft density dependence of the symmetry term are given in the upper and lower panels, respectively.
}
\label{fig_1}
\end{figure}

The predictions obtained for the elliptic flow of neutrons, protons, and 
hydrogen yields for $^{197}$Au + $^{197}$Au at 400 A MeV are shown in Fig. \ref{fig_1}.
Two values are chosen for the power-law exponent describing the density dependence 
of the potential part of the symmetry energy, $\gamma = 1.5$ (asy-stiff) and 
$\gamma = 0.5$ (asy-soft). The UrQMD outputs have been filtered with the acceptance 
of the FOPI/LAND experiment which produces the asymmetry of $v_2$ with respect to
mid-rapidity $y_{\rm lab}=0.448$. 
The neutron squeeze-out is significantly larger in the asy-stiff case 
(upper panel) than in the asy-soft case (lower panel) while the proton and hydrogen 
flows respond only weakly to the variation of $\gamma$ within the chosen interval. 

\begin{figure}[htb!]
\centering
\includegraphics*[width=70mm]{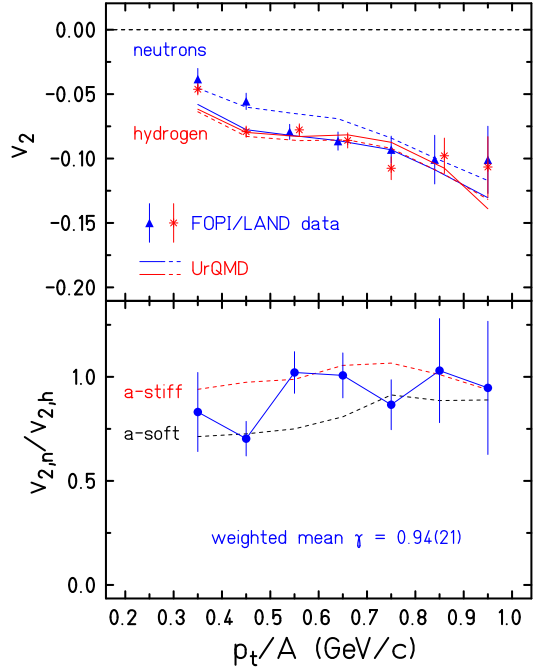}
\vskip -0.1cm

\caption{Differential elliptic flow parameters $v_2$ for neutrons (triangles) and 
hydrogens (stars, top panel) and their ratio (bottom panel) for 
central ($b<7.5$ fm) collisions of $^{197}$Au + $^{197}$Au at 400 A MeV as a function 
of the transverse momentum per nucleon $p_t/A$. The symbols represent the experimental 
data, the UrQMD predictions for $\gamma = 1.5$ (a-stiff) and $\gamma = 0.5$ (a-soft)
are given by the dashed lines.}
\label{fig_2}
\end{figure}

The comparison of the combined data set for central and mid-peripheral collisions 
with the corresponding UrQMD predictions for $b<7.5$~fm shows that the overall $p_t$ 
dependence is well described (Fig. 2, upper panel). As expected from Fig.~\ref{fig_1}, 
the squeeze-out ratio is sensitive to the stiffness of the symmetry energy (lower panel).
A linear interpolation between the predictions, averaged over  $0.3 < p_t \le 1.0$ GeV/c,
yields $\gamma = 0.94 \pm 0.21$.
A smaller but within errors consistent value $\gamma = 0.52 \pm 0.30$ is obtained
if the comparison is restricted to mid-peripheral impact-parameters 
$5.5 \le b<7.5$~fm~\cite{traut09}.
Other systematic uncertainties have been found to remain within 
$\Delta\gamma \approx 0.2$. Together with the kinetic term proportional to 
$(\rho/\rho_0)^{2/3}$, the squeeze-out data indicate a moderately soft behavior of the 
symmetry energy at supra-saturation densities. 

This result can be considered as, within errors, consistent with the density dependence 
deduced from fragmentation experiments probing nuclear matter near or below 
saturation~\cite{tsang09} and with the slightly softer density dependence resulting from 
the analysis of the pygmy dipole resonance in heavy nuclei~\cite{klimk07}. It is, 
however, inconsistent with the super-soft behavior obtained from the IBUU 
analysis~\cite{xiao09} of $\pi ^-/\pi ^+$ yield ratios reported by the FOPI collaboration 
for the same reaction $^{197}$Au + $^{197}$Au at 400 A MeV~\cite{reis07}.
According to the arguments presented in the more recent paper~\cite{zhang09}, 
the pion ratios should be sensitive to the high-density region of the reaction. 
Preliminary studies with the UrQMD model indicate that also a considerably softer symmetry 
term will be needed if the same pion ratios are to be reproduced with that model.
At present, therefore, it can only be concluded that more 
extended data sets and consistency checks in their analyses are needed in order 
to arrive at firm conclusions. In particular, it will have to be more precisely 
shown what densities have been probed, what signal modifications are to be expected 
during subsequent reaction stages, whether the deduced results are model invariant, and 
whether they are connected to the neutron richness of the studied systems. 

Illuminating discussions with M.~Di Toro and H.H. Wolter are gratefully acknowledged.

%\begin{thebibliography}{9}   % Use for  1-9  references

% For Figures insertion uncomment the next section

%\begin{figure}
%\includegraphics{figurename}
%\caption{Your caption here}
%\label{fig01} % optional figure label, must be unique
%\end{figure}

%%%%%%%%%%%%%%%%%%%%%%%%%%%%%%%%%%%%%%%%%%%%%%%%%%%%%%%%
% End of the paper
%

\begin{thebibliography}{99} % Use for 10-99 references
\itemsep -2pt 

\bibitem{lipr08}
for a recent review, see Bao-An Li, Lie-Wen Chen, Che Ming Ko, 
Phys. Rep. 464 (2008) 113.

\bibitem{klimk07}
A.~Klimkiewicz et al., Phys. Rev. C 76 (2007) 051603(R).

\bibitem{tsang09}
M.B.~Tsang et al., Phys. Rev. Lett. 102 (2009) 122701.

\bibitem{li02}
Bao-An Li, Phys. Rev. Lett. 88 (2002) 192701.

\bibitem{greco03} 
V. Greco et al., 
%V. Baran, M. Colonna, M. Di Toro, T. Gaitanos, and H.H. Wolter, 
Phys. Lett.B 562 (2003) 215. 

\bibitem{leif93}
Y. Leifels et al., Phys. Rev. Lett. 71 (1993) 963;\\
D. Lambrecht et al., Z. Phys. A 350 (1994) 115.

\bibitem{qli05}
Q. Li et al., J. Phys. G 31 (2005) 1359.

\bibitem{andro05}
A. Andronic et al., Phys. Lett. B 612 (2005) 173. 

\bibitem{traut09}
W.~Trautmann et al., Prog. Part. Nucl. Phys. 62 (2009) 425. 

\bibitem{xiao09}
Z.~Xiao et al., Phys. Rev. Lett. 102 (2009) 062502.

\bibitem{reis07}
W.~Reisdorf et al., Nucl. Phys. A 781 (2007) 459.

\bibitem{zhang09}
M.~Zhang et al., preprint arXiv:0904.0447 [nucl-th] (2009).

\end{thebibliography}
\end{document}